# Invariant Time-Dependent Exchange Perturbation Theory and its Application to the Particles Collision Problem

E.V. Orlenko[1], T. Latychevskaia[2], A.V. Evstafev[1], F.E. Orlenko[1]

[1]Theoretical Physics Department, Petersburg State Polytechnic University,

Polytechnicheskaya, 29, 195251 St.-Petersburg, Russian Federation

[2]Physics Institute, University of Zurich, Winterthurerstrasse 190, 8057 Zurich, Switzerland

Corresponding author: E. V. Orlenko, e-mail: eorlenko@mail.ru, phone: +7 (812 ) 2977871

**Abstract**

We present the formalism of Time-dependent Exchange Perturbation Theory (TDEPT) built to all orders of perturbation, for the arbitrary time dependency of perturbation. The theory takes into account the rearrangement of electrons among centres. We show how the formalism can be reduced to the standard form of invariant perturbation theory by "switching off" the re-arrangement of electrons among centres. The elements of the scattering S-matrix and transitions T-matrix and the formula for the electron scattering differential cross section are derived. The application of the theory to scattering and collision problems is discussed. As an example, we apply the theory to proton scattering on a Lithium atom, calculating the differential and total cross-sections.

# Introduction

Time-dependent perturbation theory is a well-known approach for finding an approximate solution to various problems in quantum mechanics. However, despite its powerful methods, perturbation theory cannot offer a general solution when addressing the problems of collisions between complex particles, such as molecules and atoms, especially when collision is accompanied by a change in the molecular/atomic structure. For instance, the otherwise very efficient diagram technique [1-4] fails when applied to the collision problem as it requires a wavefunction basis such that the wavefunctions of electrons assigned to different atoms are orthogonal. In turn, obtaining such an orthogonal basis given the random movement of the atomic centers is rather an irresolvable task. Overall, finding a general theoretical approach describing molecular collisions has proven to be a non-trivial problem.

Second quantization methods describing multicenter-systems have been reported previously, which are however mainly effective when addressing collisions between atoms with a many-electron structure. Within the diagram method, an attempt has been made to correct the commutation relations between the creation and annihilation operators by also taking into account permutations of electrons between orthogonal states that belong to different atomic centres. This however requires a transformation of the creation and annihilation operators and particle states into the irreducible representations of a mathematically convenient chain of subroups, which in turn leads to complicated unitary transformations.

As a result, the application of this diagram technique is limited to two-electron systems [4]. Another, more promising method employing second quantization is based on creating a so-called "chemical" Hamiltonian using a biorthogonal basis. In this method, the use of the so-called „mixed" formalism allows compressing the interactions to one- and two-center terms by applying projection operators. However, the employed biorthogonal set of spin-orbitals and the corresponding creation–annihilation operators require related unitary transformation into the "mixed" formalism and a so-called "basis extension". As a consequence, the method is effective when applied to simple systems. This explains why every given problem of a scattering event is solved by applying an ad hoc approach, as for example shown in works [5-7]. In the text-book by Davydov [2], one finds a detailed discussion problem of particles collision summarized by the conclusion that a general theoretical approach to address collision problems is required.

In this work, we attempt to develop a dedicated perturbation theory, named exchange perturbation theory (EPT), which takes into account the indistinguishability of electrons participating in multi-centre molecular / atomic collisions to any order of perturbation.

It is worth noting that the development of EPT requires careful study of the wave functions associated with electrons assigned to different atomic centres. The overlap of these wave functions is responsible for the exchange effects that play a crucial role in both adiabatic and dynamic scattering events. The exchange effects are most pronounced at the so-called intermediate atomic distances, when the molecule is re-arranging itself and its electrons are being redistributed. At these distances, the exchange effects are already larger than van der Waals forces, but at the same time, they are weak when compared to intra-atomic interactions and thus they can be considered as a perturbation. We should also mention density-functional theory (DFT) and time-dependent DFT (TDDFT) which are very popular methods for obtaining the electronic structure of many-body systems [8-11]. However, these theories are based on using approximate exchange-correlation, which is functional and limited to large molecules, such as DNA or molecular clusters, whereas, for smaller molecules, only EPT gives accurate results [12-13]. EPT, in principle, can also be applied to the calculation of large molecules or molecular clusters, but it would require significant computer time and power.

When creating a new perturbation theory formalism, one should remember that a perturbation theory should solve the two fundamental difficulties [14]:

(1) The zero-approximation antisymmetrised wave functions are non-orthogonal, and therefore they form an overfilled basis. This problem has been solved in EPT [15-18]: it has been demonstrated that the basis of non-orthogonal, antisymmetrised wavefunctions may constitute a complete set [16-18].

(2) A zero-approximation antisymmetrised wave function is not an eigenfunction of the Hamiltonian $\widehat{H}_0$ because of the non-invariance of Hamiltonian $\widehat{H}_0$ with respect to the permutations of electrons among atoms [16]. Therefore, EPT must be formulated in such a way that it would give properly symmetrised wavefunction corrections to any order of perturbation.

An overview of several EPT formalisms can be found in [14,17]. Kaplan in his book [14] classifies EPT formalisms into two groups: (1) formalisms with asymmetric unperturbed Hamiltonian $\left[\hat{A},\hat{H}_0\right]\neq 0$ and asymmetric perturbation $\left[\hat{A},\hat{V}\right]\neq 0$; (2) formalisms with symmetric zero-approximation Hamiltonian $\left[\hat{A},\hat{H}_0\right]=0$ and symmetric perturbation $\left[\hat{A},\hat{V}\right]=0$.

The first type of formalism, also named symmetry-adapted EPT formalism, has the advantage of being a simple representation of the perturbation operator, and, therefore, is often favoured as a basis for software packages. The drawback of these formalisms is that they require a *post factum* antisymmetrisation procedure, that is, after the action of the perturbation operator on the non-symmetrised wavefunction of the system: $\hat{A}\hat{V}\left(\psi^{(0)}+\psi^{(1)}+...\right)$, the problem which has been previously described in detail in [14] and references therein, and [16-18].

The second type of formalism allows the standard perturbation theory to be applied by constructing a symmetric zero-approximation Hamiltonian $\left[\hat{A},\hat{H}_0\right]=0$ with antisymmetric eigenfunction functions and a symmetrised perturbation $\left[\hat{A},\hat{V}\right]=0$ [14]. Both Hamiltonian $\widehat{H}_0$ and the perturbation operator $\hat{V}$ invariant to intercentre permutations can be obtained for the time-independent case [15-19], which allows energy corrections to be obtained using the correct antisymmetric basis of the wavefunctions. Moreover, as has been shown in [17], the energy and wavefunction corrections can be significantly simplified while preserving all the multicentre exchange inputs. Our formalism presented in this work can be assigned to the second type of formalism, for details see [17-18].

An attempt to build a formalism of time-dependent EPT (TDEPT) that accounts for the exchange of electrons among atomic centres was reported in [20]. The authors obtained series expansion of the transition amplitude up to the second order of perturbation. Examples of the implementation of TDEPT to electron and atomic scattering can be found in works [21-22].

### 1. Exchange Perturbation Theory (EPT), time-dependent perturbation

In zero-order approximation, when neglecting the interaction among particles, the coordinate part of the wavefunction of the system can be simply represented as a product of the wavefunctions of the isolated particles:

$$\Phi(r_1,...,r_N) = \prod_\alpha \psi_\alpha(r_{n_\alpha},...,r_{N_\alpha}), \qquad (1)$$

where $\alpha$ is the atomic centre number and $r_{n_\alpha},...,r_{N_\alpha}$ are the coordinates of the electrons assigned to atom $\alpha$. The spin part of the wavefunction can be expressed as a product of spinors of the same electrons: $X(\xi_1,...,\xi_N) = \prod_i \chi_i(\xi_i)$. The Hamiltonian of the system includes the kinetic energy of all electrons, the potential energy of the interaction between electrons and their atomic centres, and the interaction among electrons that belong to different centres:

$$H^0 |\Phi_n) = E_n^0 |\Phi_n), \qquad (2)$$

where $\{E_n^0\}$ are the eigenvalues of the energy and $|\Phi_n)$ are the eigenfunctions of the system and defined as a product of coordinate and spin parts. Here, the round bracket emphasizes that the eigenfunctions are not symmetrised [23-24]. We write this zero-approximation wavefunction in Dirac's symbols as vector $|\Phi^0)$.

The relatively short distances between the atomic centres cause overlap between the wavefunctions of the electrons from different centres. According to the Pauli Exclusion Principle, the complete wavefunction must be antisymmetric. An antisymmetrised wavefunction of the system can be obtained as a product of the coordinate and spin parts corresponding to conjugate Young diagrams (YD) [25,17]:

$$\Psi_n^0(r_1,...r_N,\xi_1...\xi_N) = \widehat{A}\Phi_n(r_1,...r_N) \bullet X(\xi_1...\xi_N) = \frac{1}{\sqrt{R_\lambda}} \sum_r \Phi_{nr}^{[\lambda]}(r_1,...r_N) X_{\tilde{r}}^{[\tilde{\lambda}]}(\xi_1...\xi_N),$$
(3)

where $\widehat{A}$ is the operator of antisymmetrisation, the summation is over all $r$ standard Young tableau $[\lambda]$ of a given Young diagram λ. The factor $1/\sqrt{R_\lambda}$ is the normalization factor of the Young diagram λ. $[\tilde{\lambda}]$ denotes transposed Young diagram.

The antisymmetrised vector which includes both coordinate and spin parts is now:

$$\left|\Psi_n^0\right\rangle = \frac{1}{f_n^P} \sum_p^P (-1)^{g_p} \left|\Phi_n^{0p}\right),$$
(4)

where $p$ is the permutation's number, $\frac{1}{f_n^P}$ is the normalisation factor, $g_p$ is the parity of the permutation and P is the total number of possible inter-centre permutations. The wavevector $\left|\Phi_n^{0p}\right)$ has the form provided in Eq. 3, thus including both coordinate and spin parts, and corresponds to the $p$-th permutation. We set the condition of normalisation in the form $\left(\Phi_n^{0p=0}\middle|\Psi_n^0\right\rangle = 1$, which gives the normalisation factor:

$$f_n^P = \sum_{p=0}^P (-1)^{g_p} \left(\Phi_n^{0(p=0)}\middle|\Phi_n^{0(p)}\right)$$
(5)

that is different by the factor $\sqrt{P}$ from the normalisation factor obtained in the conventional EPT where the normalisation condition is $\left\langle\Psi^0\middle|\Psi^0\right\rangle = 1$.

To find the wavefunction of the system in the presence of the time-dependent perturbation $\widehat{V}(t)$, we must solve the Schrödinger equation:

$$-\frac{\hbar}{i}\frac{\partial}{\partial t}|\Psi\rangle = \left(\widehat{H}_0 + \widehat{V}(t)\right)|\Psi\rangle.$$
(6)

for the antisymmetrised vector $|\Psi\rangle$, where the total Hamiltonian of the system $\widehat{H}_0 + \widehat{V}(t)$ is invariant. Zero-approximation Hamiltonian $\widehat{H}_0$ and time-independent perturbation operator $\widehat{V}$ were obtained in symmetrised form in works [16-18].

Because electrons which belong to different centers cannot be distinguished, Eq. 2 holds for any inter-center arrangement of the electrons, and we re-write it as

$$H^{0(p)}\left|\Phi_n^{0(p)}\right) = E_n^0\left|\Phi_n^{0(p)}\right),$$
(7)

where $\left|\Phi_n^{0(p)}\right)$ is the zero-order approximation wavefunction corresponding to the *p*-th permutation.

It has been shown that a set of eigenfunctions of the unperturbed system $\left\{\left|\Phi_n^{0(p)}\right)\right\}$ is orthogonal and complete for any permutation [16-18]:

$$\left(\Phi_m^{0(p)}\middle|\Phi_n^{0(p)}\right) = \delta_{mn};$$
$$\sum_n \left|\Phi_n^{0(p)}\right)\left(\Phi_n^{0(p)}\right| = \hat{1}; \quad (8)$$
$$\left(\Phi_m^{0(p)}\middle|\Phi_n^{0(p')}\right) \approx \delta_{mn} S_n^{(p-p')}.$$

Here $S_n^{(p-p')}$ is the overlap integral of the wavefunctions associated with the relative number of inter-center permutations of electrons *(p-p ')*. We use a "truncated" overlap which takes into account only the overlap between similar states assigned to different centers with the following hierarchy rule: $\left(\Phi_m^{0(p)}\middle|\Phi_n^{0(p')}\right) << \left(\Phi_n^{0(p)}\middle|\Phi_n^{0(p')}\right) \sim \left(\Phi_m^{0(p)}\middle|\Phi_m^{0(p')}\right)$.

As has been shown in [16-18], the basis of antisymmetric functions of the unperturbed system is complete:

$$\sum_n \left|\Psi_n^0\right\rangle\frac{f_n^P}{P}\left(\Phi^{0(0)}_n\right| = \hat{1}, \quad (9)$$

where antisymmetrised vector $\left|\Psi_n^0\right\rangle$ is given by Eq. 4.

To write Hamiltonian and perturbation operators in invariant form we introduce the projection operator:

$$\Lambda^{(p)} = \sum_n \left|\Phi_n^{0(p)}\right)\frac{f_n}{P}\left(\Phi_n^{0(p)}\right|, \quad (10)$$

which, when applied to the wavevector $\left|\Psi_i^0\right\rangle$ given by Eq. 4 at *n=i*, selects the wavefunction corresponding to *p*-th permutation: $\Lambda^{(p)}\left|\Psi_i^0\right\rangle = (-1)^{g_p}\left|\Phi_i^{0(p)}\right)$ [18].

We use operator $\Lambda^{(p)}$ to write the Hamiltonian and perturbation operators in a form invariant under inter-centre permutations:

$$\hat{V}(t) = \sum_{p=0}^P V^{(p)}(t)\Lambda^{(p)}, \quad \hat{H}_0 = \sum_{p=0}^P H^{0(p)}\Lambda^{(p)}, \quad (11)$$

here $H^{0(p)}$ and $V^{(p)}(t)$ are the Hamiltonian and perturbation operators corresponding to the *p*-th permutation of the electrons between the centers. The antisymmetrised zero-approximation wavevector $\left|\Psi_i^0\right\rangle$ is an eigenvector of the invariant Hamiltonian of the unperturbed system: $\hat{H}_0\left|\Psi_i^0\right\rangle = E_i^0\left|\Psi_i^0\right\rangle$, where energy eigenvalues $E_i^0$ are real [15,19,18]. Note, that through the entire text we use the "hat" symbol for symmetrised operators and do not use the "hat" symbols for operators that are related to *p*-th permutation.

We search for the wavevector $\left|\Psi\right\rangle$ which is the solution to the Schrödinger equation, Eq. 6, by the method of successive approximations, as it is typically solved in conventional perturbation theory.

## 2. Perturbation theory to the first order

Let $\left|\Psi^1(t)\right\rangle$ be the first-order correction to the wavefunction $\left|\Psi\right\rangle$, which in the first-order approximation satisfies the following equation

$$-\frac{\hbar}{i}\left|\dot{\Psi}^1(t)\right\rangle = \hat{H}_0\left|\Psi^1(t)\right\rangle + \hat{V}(t)\left|\Psi_i^0\right\rangle. \tag{12}$$

We write the solution to this equation in the form of series

$$\left|\Psi^1(t)\right\rangle = \sum_n C_n^{(1)}(t)\exp\left(-\frac{i}{\hbar}E_n t\right)\left|\Psi_n^0\right\rangle, \tag{13}$$

where the expansion coefficients $C_n^{(1)}(t)$ are to be found.

Next, we introduce two skew-projection operators $\hat{P}_i$ and $\hat{O}_i$. The skew-projection operator $\hat{P}_i = \left|\Psi_i^0\right\rangle\left(\Phi_i^{0(0)}\right|$ projects onto the subspace of the vectors parallel to the vector $\left|\Psi_i^0\right\rangle$ giving $\hat{P}_i\left|\Psi_i^0\right\rangle \equiv \left|\Psi_i^0\right\rangle$. The skew-projection operator $\hat{O}_i = 1 - \hat{P}_i$ projects onto the subspace of vectors orthogonal to the vector $\left|\Psi_i^0\right\rangle$ giving $\hat{O}_i\left|\Psi_i^0\right\rangle \equiv 0$.

By substitution Eq. 13 into Eq. 12 and applying operator $\hat{O}_i$ to both sides of the equation, we obtain the following expression for coefficients $\dot{C}_n^{(1)}(t)$ (see Appendix I for details):

$$-\frac{\hbar}{i}\dot{C}_n^{(1)}(t) = \frac{f_0}{P}\exp(-i\omega_{ni}t)\left(\Phi_n^{0(0)}\left|\hat{O}_i\hat{V}\right|\Psi_i^0\right), \tag{14}$$

where we used the property of completeness of the antisymmetrised basis given by Eq. 9 and introduced frequencies: $\frac{1}{\hbar}(E_i - E_n) = \omega_{in}$. The solution to Eq. 14 can be written in the form of a definite integral:

$$C_n^{(1)}(t) = \frac{f_0}{i\hbar P}\int_0^t \exp(i\omega_{ni}t')\left(\Phi_n^{0(0)}\left|\hat{O}_i\hat{V}\right|\Psi_i^0\right)dt'. \tag{15}$$

Next, we use the following results

$$\hat{O}_i\left|\Psi_n^0\right\rangle \approx \left|\Psi_n^0\right\rangle, \tag{16}$$

$$(\Phi_n^{0(0)}|\hat{O}_i = (\Phi_n^{0(0)}| - (\Phi_n^{0(0)}|\Psi_i^0\rangle(\Phi_i^{0(0)}| \approx (\Phi_n^{0(0)}|,$$

$$(\Phi_i^{0(0)}|\hat{O}_i = (\Phi_i^{0(0)}| - (\Phi_i^{0(0)}|\Psi_i^0\rangle(\Phi_i^{0(0)}| \equiv 0$$

which were derived in [16], and we obtain the first-order corrections to the expansion coefficients as

$$C_n^{(1)}(t) = \frac{f_n}{i\hbar P}\int_0^t \exp(i\omega_{ni}t')\left(\Phi_n^{0(0)}\left|\hat{V}\right|\Psi_i^0\right)dt'. \tag{17}$$

Equation 17 describes the amplitude of the transition between states $i$ and $n$. By substitution $C_n^{(1)}(t)$ from Eq. 17 into Eq. 13 we find the first-order correction to the wavevector of the system.

The matrix elements of the perturbation operator in Eq. 17 can be rewritten using the perturbation operator expressed through projection operator $\Lambda^{(p)}$ see Eq. 11:

$$\left(\Phi_n^{0(0)}\left|\hat{V}\right|\Psi_i^0\right) = \left(\Phi_n^{0(0)}\left|\sum_{p=0}^{P} V_p \frac{(-1)^{g_p}}{f_i^P}\right|\Phi_i^{0(p)}\right) = \frac{f_n^P}{f_i^P}\left\langle\Psi_n^0\left|V_{p=0}\right|\Phi_i^{0(0)}\right). \tag{18}$$

By substitution the result of Eq. 18 into Eq. 17 we obtain the final expression for the coefficients:

$$C_n^{(1)}(t) = \frac{(f_n)^2}{i\hbar P}\int_0^t \exp(i\omega_{ni}t')\left\langle\Psi_n^0\left|V_{p=0}\right|\Phi_i^{0(0)}\right)dt'. \tag{19}$$

If we neglect the exchange interaction, and replace $\Psi \rightarrow \Phi$, Eq. 19 will give the same result as the result obtained in conventional perturbation theory.

### 3. Perturbation theory to the second and higher orders

The time-dependent Schrödinger equation, Eq. 6, is now written up to the second order. By substituting the wavevector found to the first order and after some simplifications, we obtain the equation for the second-order correction to the wavefunction:

$$-\frac{\hbar}{i}\left|\dot{\Psi}_i^2\right\rangle = \hat{H}_0\left|\Psi_i^2\right\rangle + \hat{V}\left|\Psi_i^1\right\rangle. \tag{20}$$

As above, we seek the solution in the form of series:

$$\left|\Psi_i^2\right\rangle = \sum_n{'} C_n^{(2)}(t)\exp\left(-\frac{i}{\hbar}E_n t\right)\left|\Psi_n^0\right\rangle. \tag{21}$$

where the expansion coefficients $C_n^{(2)}(t)$ are to be found.

The procedure for finding the coefficients is similar to the procedure described above. We obtain:

$$-\frac{\hbar}{i}\sum_n{'} P\dot{C}_n^{(2)}\exp\left(-\frac{i}{\hbar}E_n t\right)\left|\Psi_n^0\right\rangle = \hat{O}_i\hat{V}\left|\Psi_i^1\right\rangle. \tag{22}$$

Next, we can use the property of completeness Eq. 9 and re-write the right part of Eq. 22 as

$$\hat{O}_i\hat{V}\left|\Psi_i^1\right\rangle = \sum_n{'} \frac{f_n}{P}\left|\Psi_n^0\right\rangle\left(\Phi_n^{0(0)}\left|\hat{O}_i\hat{V}\right|\Psi_i^1\right\rangle. \tag{23}$$

Substituting Eq. 23 into Eq. 22, after some re-arrangement of the terms, we obtain:

$$C_n^{(2)}(t) = \frac{f_b}{i\hbar P}\int_0^t dt'\exp(iE_n t'/\hbar)\left(\Phi_n^{0(0)}\left|\hat{O}_i\hat{V}\right|\Psi_i^1(t')\right\rangle.$$

Applying the result of Eq. 18 we re-write the last equation as

$$C_{n_2}^{(2)}(t) = \frac{f_b}{i\hbar P}\int_0^t dt'\exp(iE_{n_2}t'/\hbar)\left(\Phi_{n_2}^{0(0)}\left|\widehat{V}\right|\Psi_i^1(t')\right) =$$

$$= \left(\frac{f_0^2}{i\hbar P}\right)^2 \sum_{n_1}\int_0^t dt'\exp(i\omega_{n_2 n_1}t')\langle\Psi_{n_2}^0|V_{p=0}|\Phi_{n_1}^{0(0)}\rangle\int_0^{t'} dt''\langle\Psi_{n_1}^0|V_{p=0}|\Phi_i^{0(0)}\rangle\exp(i\omega_{n_1 i}t'').$$

(24)

This expression is almost the same as that in conventional perturbation theory with the orthogonal basis of eigenfunctions. The difference is that Eq. 24 includes matrix elements enclosing overlap integrals produced by accounting for exchange and superexchange contributions.

The second-order correction to the wavevector in its full form is:

$$|\Psi_i^2\rangle = \sum_{n_2}{}' \exp\left(-\frac{i}{\hbar}E_{n_2}t\right)|\Psi_{n_2}^0\rangle \times$$

$$\times\left(\frac{f_0^2}{i\hbar P}\right)^2 \sum_{n_1}\int_0^t dt'\exp(i\omega_{n_2 n_1}t')\langle\Psi_{n_2}^0|V_{p=0}(t')|\Phi_{n_1}^{0(0)}\rangle\int_0^{t'} dt''\langle\Psi_{n_1}^0|V_{p=0}(t'')|\Phi_i^{0(0)}\rangle\exp(i\omega_{n_1 i}t'').$$

(25)

From the expressions for the first-order and second order corrections, $|\Psi_i^1\rangle$ and $|\Psi_i^2\rangle$, respectively, we can see the general tendency and derive the following expression for an *n*-th correction to the wave function:

$$|\Psi_i^n\rangle = \sum_{n_n}{}'\sum_{n_{n-1}}{}'\ldots\sum_{n_1}{}'\left(\frac{f_0^2}{i\hbar P}\right)^n|\Psi_{n_n}^0\rangle\exp\left(-\frac{i}{\hbar}E_{n_n}t\right)\times$$

$$\int_0^t dt_1\langle\Psi_{n_n}^0|V_{p=0}(t_1)|\Phi_{n_{n-1}}^{0(0)}\rangle\exp(i(\omega_{n_n n_{n-1}}t_1)\int_0^{t_1} dt_2\langle\Psi_{n_{n-1}}^0|V_{p=0}(t_2)|\Phi_{n_{n-2}}^{0(0)}\rangle\exp(i\omega_{n_{n-1}n_{n-2}}t_2)\ldots \quad (26)$$

$$\ldots\int_0^{t_{n-1}} dt_n\langle\Psi_{n_1}^0|V_{p=0}(t_n)|\Phi_i^{0(0)}\rangle\exp(i\omega_{n_1 i}t_n).$$

A note to this equation: the transition matrix elements related to later times are placed before the transition matrix elements related to earlier times, as it holds:

$$t > t_1 > t_2 > \ldots > t_n > 0.$$

Equation 26 can be transformed to a more symmetric form by introducing the time ordering operator $\hat{\tau}$:

$$|\Psi_i^n\rangle = \frac{1}{n!}\left(\frac{f_0^2}{i\hbar P}\right)^n \sum_{n_n}{}'\sum_{n_{n-1}}{}'\ldots\sum_{n_1}{}'|\Psi_{n_n}^0\rangle\exp\left(-\frac{i}{\hbar}E_{n_n}t\right)\times$$

$$\times\int_0^t\ldots\int_0^t \hat{\tau}\{\langle\Psi_{n_n}^0|V_{p=0}(t_1)|\Phi_{n_{n-1}}^{0(0)}\rangle\exp(i(\omega_{n_n n_{n-1}}t_1)\ldots \quad (27)$$

$$\ldots\langle\Psi_{n_1}^0|V_{p=0}(t_n)|\Phi_i^{0(0)}\rangle\exp(i\omega_{n_1 i}t_n)\}dt_1\ldots dt_n.$$

This expression defines the corrections to the wavefunction of a multi-centered system with exchange effects to any order of perturbation. Equation 27 can also be applied to the system where the inter-center exchange is negligible, in which case Eq. 27 transforms into the expression for wavefunction corrections obtained in conventional time-dependent perturbation theory.

The coefficient corresponding to the transition from the initial state $|i\rangle$ to the final state $|f\rangle$ can be written in an invariant form:

$$C_f(t) = \langle \Psi_f^0 | \hat{\tau} \exp(-\frac{if_0^2}{\hbar P}\int_0^t \widehat{W}(t')dt') | \Phi_i^{0(0)} \rangle,  \qquad (28)$$

where we introduced operator

$$\widehat{W}(t) = e^{\frac{i}{\hbar}H_0^{p=0}t} V^{p=0}(t) e^{-\frac{i}{\hbar}H_0^{p=0}t}. \qquad (29)$$

The probability of transition at the time $t$ is given by $w_{fi}(t) = |C_f(t)|^2$ when transition occurs between states of a discrete spectrum, and it is given by $dw_{fi}(t) = |C_f(t)|^2 dv_f$ when the transition occurs in the range $dv_f$ of continuous spectrum.

## 4. S-scattering and T-matrix elements

We consider two non-interacting sub-systems (examples include: an electron and an atom or an alpha-particle, an ion and an atom, two atoms, an atom and a molecule, two molecules, etc., also possible are a light wave and a molecule which decomposes into fragments, and similar). Two states $|i\rangle$, $|f\rangle$ and their energies $E_i, E_f$ are the solution to the equation $\widehat{H}_0 |\Psi_i^0\rangle = E_i^0 |\Psi_i^0\rangle$ or eigenstates and eigenenergies of the whole system Hamiltonian $\widehat{H}_0$. Perturbation which causes interaction between the two subsystem and a transition between the two states is described by operator $\widehat{V} = \sum_{p=0}^{P} V^{(p)} \Lambda^{(p)}$, which is time-independent in the Schrödinger representation.

The time-dependent transition coefficient $C_f(t) \equiv C_{if}(t)$ given by Eq. 28 turns into a scattering S-matrix element when the time variable $t$ ranges from $-\infty$ to $+\infty$:

$$C_{if}(\infty) = \langle \Psi_f^0 | S | \Phi_i^{0(0)} \rangle = \langle \Psi_f^0 | \hat{\tau} \exp(-\frac{i}{\hbar}\frac{f_0^2}{P}\int_{-\infty}^{\infty} \widehat{W}(t)dt) | \Phi_i^{0(0)} \rangle \equiv$$

$$\equiv \langle \Psi_f^0 | \{1 + \frac{1}{i\hbar}\frac{f_0^2}{P}\int_{-\infty}^{\infty} \widehat{W}(t)dt) + \frac{1}{(i\hbar)^2}\left(\frac{f_0^2}{P}\right)^2 \int_{-\infty}^{\infty} dt_1 \int_{-\infty}^{\infty} dt_2 \widehat{W}(t_1)\widehat{W}(t_2)) + ...\} | \Phi_i^{0(0)} \rangle =$$

$$= \sum_{\alpha}^{\infty} \langle \Psi_f^0 | S^{(\alpha)} | \Phi_i^{0(0)} \rangle. \qquad (30)$$

Using this result, we obtain the first-order correction to the S-matrix elements:

$$\langle\Psi_f^0|S^{(1)}|\Phi_i^{0(0)}\rangle = -\frac{i}{\hbar}\frac{f_0^2}{P}\langle\Psi_f^0|V_0|\Phi_i^{0(0)}\rangle\int_{-\infty}^{\infty}e^{i(E_f-E_i)\frac{t}{\hbar}}dt =$$
$$= -2\pi i\delta(E_f - E_i)\frac{f_0^2}{P}\langle\Psi_f^0|V_0|\Phi_i^{0(0)}\rangle, \quad (31)$$

where we took into account the formula for the time-dependent transition coefficient given by Eq. 28.

The second-order correction to the S-matrix element equals to [2]:

$$\langle\Psi_f^0|S^{(2)}|\Phi_i^{0(0)}\rangle = -2\pi i\delta(E_i - E_f)\left(\frac{f_0^2}{P}\right)^2\sum_n\frac{\langle\Psi_f^0|V_0|\Phi_n^{0(0)}\rangle\langle\Psi_n^0|V_0|\Phi_i^{0(0)}\rangle}{E_i - E_n + i\eta}. \quad (32)$$

Other higher-order corrections to the S-matrix elements can be found in a similar manner.

Because the interaction between the atomic centers is much slower than the interaction between their electrons, adiabatic approximation can be applied and interaction between atoms can be considered as a time-independent perturbation. The transitions caused by that perturbation are transitions between an initial state and a different final state, so that $\langle\Psi_f|\Phi_i\rangle \approx 0$ is fulfilled. Thus, we can rewrite matrix elements of the S-matrix in Eq.30 in the form

$$\langle\Psi_f^0|S|\Phi_i^{0(0)}\rangle = -2\pi i\delta(E_f - E_i)\langle\Psi_f^0|\hat{T}|\Phi_i^{0(0)}\rangle, \quad (33)$$

where we introduced an operator of transition on the energy surface $\hat{T}$ and the result is expressed through its matrix elements $\langle\Psi_f^0|\hat{T}|\Phi_i^{0(0)}\rangle$. We would like to emphasize that matrix elements of operator $\hat{T}$, obtained here, unlike the commonly used matrix elements of operator $\hat{T}$ [2], account for all possible electron permutations between the two subsystem, which was achieved by antisymmetrisation of the wavefunction of the whole system.

The matrix elements of operator $\hat{T}$ can be expressed as:

$$\langle\Psi_f^0|\hat{T}|\Phi_i^{0(0)}\rangle = \frac{f_0^2}{P}\langle\Psi_f^0|V_0|\Phi_i^{0(0)}\rangle + \left(\frac{f_0^2}{P}\right)^2\sum_n\frac{\langle\Psi_f^0|V_0|\Phi_n^{0(0)}\rangle\langle\Psi_n^0|V_0|\Phi_i^{0(0)}\rangle}{E_i - E_n + i\eta} + ... +$$
$$+ \left(\frac{f_0^2}{P}\right)^\nu\sum_{n,n_1...n_{\nu-2}}\frac{\langle\Psi_f^0|V_0|\Phi_n^{0(0)}\rangle...\langle\Psi_{n_{\nu-2}}^0|V_0|\Phi_i^{0(0)}\rangle}{(E_i - E_n + i\eta)(E_i - E_{n_1} + i\eta)...(E_i - E_{n_{\nu-2}} + i\eta)}... \quad (34)$$

where ν is the expansion order. The matrix elements of various orders contained in Eq. 34 can be displayed graphically using Feynman diagrams, as has been demonstrated in previous works [19,18].

Because the functions $\Phi_n^{0(0)}$ are the eigenfunctions of the non-symmetrical Hamiltonian $H_{(p=0)}^0$, the second term on the right side of Eq. 34 can be re-written as:

$$\left(\frac{f_0^2}{P}\right)^2 \sum_n \frac{\langle \Psi_f^0 |V_0| \Phi_n^{0(0)} \rangle \langle \Psi_n^0 |V_0| \Phi_i^{0(0)} \rangle}{E_i - E_n + i\eta} =$$
$$= \left(\frac{f_0^2}{P}\right)^2 \langle \Psi_f^0 |V_0| \Phi_n^{0(0)} \rangle \left(\frac{f_0^2}{P}\right)^{-1} \langle \Psi_n^0 | \left(E_i - H_{p=0}^0 + i\eta\right)^{-1} |\Phi_n^{0(0)} \rangle \langle \Psi_n^0 |V_0| \Phi_i^{0(0)} \rangle, \quad (35)$$

where we applied the property of completeness, Eq. 9.

By substituting Eq. 35 into Eq. 34, we obtain the following expression for operator $\hat{T}$:

$$\hat{T} = \left(\frac{f_0^2}{P}\right) V_0 + \left(\frac{f_0^2}{P}\right) V_0 \left(\frac{f_0^2}{P}\right)^{-1} \left(E_i - H_{p=0}^0 + i\eta\right)^{-1} \left(\frac{f_0^2}{P}\right) V_0 +$$
$$+ \left(\frac{f_0^2}{P}\right) V_0 \left(\frac{f_0^2}{P}\right)^{-1} \left(E_i - H_{p=0}^0 + i\eta\right)^{-1} \left(\frac{f_0^2}{P}\right) V_0 \left(\frac{f_0^2}{P}\right)^{-1} \left(E_i - H_{p=0}^0 + i\eta\right)^{-1} \left(\frac{f_0^2}{P}\right) V_0 + \ldots$$
$$(36)$$

T-matrix can be found as a solution to Eq. 36 by successive approximations to the following operator equation:

$$\hat{T} = V_0^{\mathbb{N}} + V_0^{\mathbb{N}} \left(\frac{f_0^2}{P}\right)^{-1} \left(E_i - H_{p=0}^0 + i\eta\right)^{-1} \hat{T}, \quad (37)$$

where $V_0^{\mathbb{N}} = \left(\frac{f_0^2}{P}\right) V_0$ is a renormalized perturbation operator.

The transition probability per unit of time is given by

$$w_{fi} = \frac{\left|\langle \Psi_f^0 |S| \Phi_i^{0(0)} \rangle\right|^2}{\lim_{\tau \to \infty} \int_{-\tau}^{\tau} dt} = \frac{2\pi}{\hbar} \delta(E_f - E_i) \left|\langle \Psi_f^0 |\hat{T}| \Phi_i^{0(0)} \rangle\right|^2, \quad (38)$$

where we used Eq. 33.

To obtain an expression for the cross section of scattering events and reactions, we divide the result of Eq. 38 by the flux-density of incident particles: $j_i = \hbar k_i / \mu_i$, where $k_i$ is the wavevector of the relative movement of the incident particles and $\mu_i$ is their reduced mass:

$$\sigma_{fi} = \frac{2\pi \mu_i}{\hbar^2 k_i} \delta(E_f - E_i) \left|\langle \Psi_f^0 |\hat{T}| \Phi_i^{0(0)} \rangle\right|^2. \quad (39)$$

When the final state is within the continuous spectrum, the transition probability given by Eq. 38 must be multiplied by the number of final states in the volume per unit energy interval $\rho(E_f)$:

$$w_{fi} = \frac{2\pi}{\hbar}\delta(E_f - E_i)\left|\langle\Psi_f^0|\hat{T}|\Phi_i^0\rangle\right|^2 \rho(E_f). \tag{40}$$

By integration over all the possible energy of the final states, the formula for the probability of a scattering event per unit of time given by Eq. 40 can be transformed into an expression which accounts for all inter-center exchange contributions.

### 5. Collisions with exchange of electrons

The strength of the interaction between particles during collisions is described by a scattering cross-section, or a differential cross-section. The differential scattering cross section is obtained by substituting the given $\rho(E_f)$ into Eq. 40 and dividing the result by the flux-density of the incident particles $j_i$. In this section, we derive the expression for $\rho(E_f)$, and consequently, for the differential cross section.

In the following, we consider the collision associated with the redistribution of electrons, as for example the collisions of positive ions with neutral atoms accompanied by charge transfer. The initial Hamiltonian of the system corresponds to the initial distribution of the electrons between the centers (permutation $p = 0$):

$$H_0^{p=0} = -\frac{\hbar^2}{2\mu_i}\nabla_i^2 + H_i^{p=0}(r_1, r_2, ...)_i, . \tag{41}$$

which is the sum of the operator describing the kinetic energy of the relative motion (with reduced mass) and the operator of internal state of the colliding atoms (molecules), index $i$ denotes the initial state.

For the initial state, the energy eigenvalues $E_i$ and eigenvector $|\Phi_i\rangle$, accounting for spin distribution, amounts to

$$E_i = -\frac{\hbar^2}{2\mu_i}k_i^2 + \varepsilon_{n_i} \tag{42}$$

$$|\Phi_i\rangle = \Phi_i(\vec{R}-\vec{r}_1, \vec{R}-\vec{r}_2, \vec{R}-\vec{r}_3...)\chi(\xi_1, \xi_2; \xi_3...)\exp(i\vec{k}_i\vec{R}), \tag{43}$$

where $\vec{R}$ is vector originating at one colliding particle and pointing to the other colliding particle, its absolute value $R$ is the distance between colliding particles.

As introduced above, see Eq. 11, the symmetric Hamiltonian form describing the final state can be written as

$$\hat{H}_{0f} = \sum_{p=0}^{P} H_f^{0(p)}\Lambda_f^{(p)}$$

$$H_f^{0(p)} = -\frac{\hbar^2}{2\mu_f}\nabla_f^2 + H_f^p(r_1, r_2, ...)_f \tag{44}$$

The energy eigenvalues $E_f$ and antisymmetric eigenvector of $\hat{H}_{0f}$, accounting for the permutations, are given by

$$E_f = -\frac{\hbar^2}{2\mu_i}k_f^2 + \varepsilon_{n_f} \tag{45}$$

$$|\Psi_f\rangle = \hat{A}\Phi_f(\vec{R}-\vec{r}_1, \vec{R}-\vec{r}_2, \vec{R}-\vec{r}_3...)\chi(\xi_1, \xi_2; \xi_3...)\exp(i\vec{k}_f\vec{R}). \tag{46}$$

The number of the final states per unit energy interval for scattering in the direction of the unit vector $\vec{n}_f$ in the element of solid angle $d\Omega$ is given by the expression: $d\rho(E_f) = \dfrac{\mu_f k_f}{(2\pi)^3 \hbar^2} d\Omega$. By substituting $d\rho(E_f)$ into Eq. 40 we obtain:

$$\frac{d\sigma_{fi}}{d\Omega} = j^{-1}\frac{dw_{fi}}{d\Omega} = \frac{\mu_i \mu_f k_f}{(2\pi\hbar^2)^2 k_i}\left|\langle\Psi_f^0|\hat{T}|\Phi_i^0\rangle\right|^2. \tag{47}$$

The transition operator $\hat{T}$ given by Eq. 37 can be re-written in the form

$$\hat{T} = V_0^{\mathbb{N}} + V_0^{\mathbb{N}}\left(\frac{f_0^2}{P}\right)^{-1}(E_i - H + i\eta)^{-1}V_0^{\mathbb{N}}, \tag{48}$$

where $H$ is the total Hamiltonian of the system, see Appendix II.

### 6. Scattering of proton by Lithium atom with electron exchange

In this section we demonstrate how the obtained relations can be applied to a practical problem of proton scattering on Lithium atom: $\mathrm{Li} + p \rightarrow \mathrm{Li}^+ + \mathrm{H}$. $\vec{R}$ is vector originating at proton and pointing to Lithium atom. Figure 1 illustrates the arrangement of the particles.

As the initial permutation ($p = 0$) we assume the following arrangement: three electrons of the Lithium atom and the incident proton. We label the electrons by 1, 2 and 3.

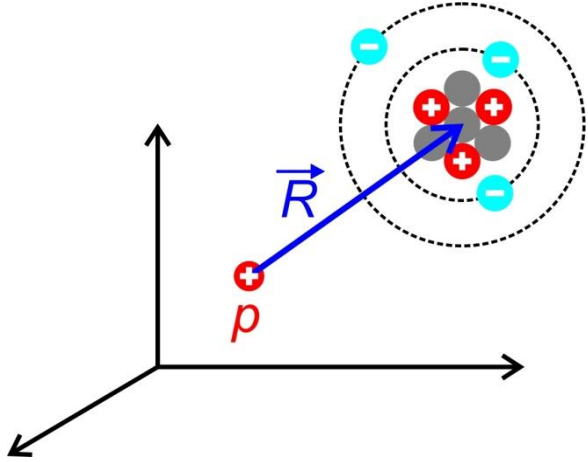

Fig. 1. Arrangement of the particles in the proton–Lithium atom collision.

The relative motion of the proton and the atom is determined by the operator of kinetic energy $\left(-\dfrac{\hbar^2}{2\mu}\nabla^2_{p-\mathrm{Li}}\right)$.

Interaction between proton and the atom is determined by the operator:

$$V_0 \equiv V(1,2,3) = -\frac{e^2}{r_{p1}} - \frac{e^2}{r_{p2}} - \frac{e^2}{r_{p3}} + \frac{3e^2}{R}. \tag{49}$$

The motion of electrons is described by the Hamiltonian $H(1,2,3) = -\frac{\hbar^2}{2m}(\nabla_1^2 + \nabla_2^2 + \nabla_3^2) - \frac{Ze^2}{r_{Li1}} - \frac{Ze^2}{r_{Li2}} - \frac{Ze^2}{r_{Li3}}$ whose antisymmetric eigenfunction with respect to intra-atomic electron permutations is given by

$$\Phi_{Li}(\vec{R}-\vec{r}_1, \vec{R}-\vec{r}_2, \vec{R}-\vec{r}_3) \otimes \chi(\xi_1, \xi_2; \xi_3) = \frac{1}{f_{Li}}\left(\Psi_{Li1}(\vec{r}_1, \vec{r}_2, \vec{r}_3)X_{Li1}(1,2,3) + \Psi_{Li2}(\vec{r}_1, \vec{r}_2, \vec{r}_3)X_{Li2}(1,2,3)\right)$$

(50)

and corresponds to the energy eigenvalues $\varepsilon_n$. Antisymmetrisation of the atomic wavefunction given by Eq. 50 over intratomic electrons permutations is performed using 4 independent Young's operators corresponding to the Young diagram depicted in Fig. 1: $\omega_{11}^{[21]}; \omega_{12}^{[21]}; \omega_{21}^{[21]}; \omega_{22}^{[21]}$. Here the superscripts [21] describe the ordering of empty boxes in the Young diagram: two boxes in the upper row and one box in the bottom row, and the subscripts describe the filling of the boxes [25]. For example,

$$\omega_{11}^{[21]} = \frac{1}{\sqrt{12}}(2 + 2P_{12} - P_{23} - P_{13} - P_{123} - P_{132}) \quad \text{and} \quad \omega_{12}^{[21]} = (P_{23} - P_{13} - P_{123} + P_{132}).$$

$P_{12}$ denotes the permutation of electrons 1 and 2; $P_{123}$ denotes the cyclic permutation of electrons 1, 2 and 3.

The components of the spatial parts of the eigenfunction in Eq. 50 are given by

$$\Psi_{Li1}(\vec{r}_1, \vec{r}_2, \vec{r}_3) = \omega_{11}^{[21]}\phi_{1s}(\vec{R}-\vec{r}_1)\phi_{1s}(\vec{R}-\vec{r}_2)\varphi_{2s}(\vec{R}-\vec{r}_3);$$
$$\Psi_{Li2}(\vec{r}_1, \vec{r}_2, \vec{r}_3) = \omega_{12}^{[21]}\phi_{1s}(\vec{R}-\vec{r}_1)\phi_{1s}(\vec{R}-\vec{r}_2)\varphi_{2s}(\vec{R}-\vec{r}_3) = 0$$

(51)

and the corresponding spin parts are

$$X_{Li1} = \omega_{22}^{[21]}\chi(123) = \omega_{22}^{[21]}|\alpha 1\beta 2\alpha 3\rangle = \frac{1}{\sqrt{12}}\{2|\alpha 1\beta 2\alpha 3\rangle + |\alpha 1\alpha 2\beta 3\rangle + |\alpha 1\beta 2\alpha 3\rangle - 2|\beta 1\alpha 2\alpha 3\rangle -$$
$$-|\alpha 1\alpha 2\beta 3\rangle - |\beta 1\alpha 2\alpha 3\rangle\} = \frac{\sqrt{3}}{2}\{|\alpha 1\beta 2\alpha 3\rangle - |\beta 1\alpha 2\alpha 3\rangle\}$$
$$X_{Li2} = \omega_{21}^{[21]}\chi(123) = \omega_{21}^{[21]}|\alpha 1\beta 2\alpha 3\rangle = \frac{1}{2}\{-|\alpha 1\alpha 2\beta 3\rangle + |\alpha 1\beta 2\alpha 3\rangle - |\alpha 1\alpha 2\beta 3\rangle + |\beta 1\alpha 2\alpha 3\rangle\} =$$
$$= \frac{1}{2}\{-2|\alpha 1\alpha 2\beta 3\rangle + |\alpha 1\beta 2\alpha 3\rangle + |\beta 1\alpha 2\alpha 3\rangle\}.$$

(52)

The Hamiltonian of the unperturbed system corresponding to the initial permutation of electrons is $H_0^{p=0} = H(1,2,3) - \frac{\hbar^2}{2\mu}\nabla_{p-Li}^2$.

In the initial state, the motion of the proton and the Lithium atom relative to the center of mass is described as a plane wave with the wavevector of relative motion $\vec{k}_i$. Similarly, in the final state, the motion of the Hydrogen atom and the Lithium ion is described as a plane wave with the wavevector of relative motion $\vec{k}_f$.

The initial state is described by the function

$$|\Phi_i\rangle = \Phi_{Li}(\vec{R}-\vec{r}_1, \vec{R}-\vec{r}_2, \vec{R}-\vec{r}_3) \otimes \chi(\xi_1, \xi_2; \xi_3)\exp(i\vec{k}_i\vec{R}) =$$
$$= \frac{1}{f_{Li}}\left(\Psi_{Li1}(\vec{r}_1, \vec{r}_2, \vec{r}_3)X_{Li1}(1,2,3) + \Psi_{Li2}(\vec{r}_1, \vec{r}_2, \vec{r}_3)X_{Li2}(1,2,3)\right)\exp(i\vec{k}_i\vec{R}).$$

(53)

Here, the factor $f_{Li}$ is found from the normalization condition, which, after some calculations, results in

$$f_{Li}^2 = \left\{\langle\Psi_{Li1}(\vec{r}_1,\vec{r}_2,\vec{r}_3)|\Psi_{Li1}(\vec{r}_1,\vec{r}_2,\vec{r}_3)\rangle\langle X_{Li1}|X_{Li1}\rangle + \langle\Psi_{Li2}(\vec{r}_1,\vec{r}_2,\vec{r}_3)|\Psi_{Li2}(\vec{r}_1,\vec{r}_2,\vec{r}_3)\rangle\langle X_{Li2}|X_{Li2}\rangle\right\} = 4 \quad (53a)$$

and thus $f_{Li}=2$.

The antisymmetric vector of the final state is obtained by applying the normalized Young operator [25] to the wave function of the Hydrogen–Lithium-ion system:

$$|\Phi_f\rangle = \Phi_{Li^+}(\vec{R}_f-\vec{r}_1,\vec{R}_f-\vec{r}_2)\Phi_H(\vec{r}_3)\chi(\xi_1,\xi_2;\xi_3)\exp(i\vec{k}_f\vec{R}_f) \quad (54)$$

and (for the open channel $\frac{\hbar^2 k_f^2}{2\mu} = \varepsilon_i - \varepsilon_n + \frac{\hbar^2 k_i^2}{2\mu} \geq 0$ ) is given by

$$|\Psi_f\rangle = \frac{1}{f_0}\left(\Psi_1(\vec{r}_1,\vec{r}_2,\vec{r}_3)X_1(1,2,3) + \Psi_2(\vec{r}_1,\vec{r}_2,\vec{r}_3)X_2(1,2,3)\right)\exp(i\vec{k}_f\vec{R}). \quad (55)$$

For a three-electron system, where two electrons both occupy the 1s² orbital in a singlet state, there is only one Young diagram possible, as depicted in Fig. 2(a). This Young diagram allows the two possible combinations shown in Fig. 2(b) and (c).

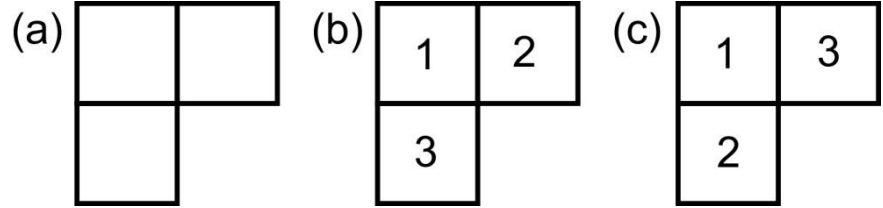

Fig. 2. Young diagrams for three-electron system. (a) Young diagram. (b) and (c) filled Young tableaux with two possible combinations.

After applying the Young operators to the functions described by Eq. 54, the spatial parts of the antisymmetric wavefunction of Hydrogen atom–Lithium ion system can be expressed as:

$$\Psi_1(\vec{r}_1,\vec{r}_2;\vec{r}_3) = \omega_{11}^{[21]}\Phi_{Li^+}(\vec{R}-\vec{r}_1,\vec{R}-\vec{r}_2)\Phi_H(\vec{r}_3) =$$
$$= \frac{1}{\sqrt{12}}(2 + 2P_{12} - P_{23} - P_{13} - P_{123} - P_{132})\Phi_{Li^+}(\vec{R}-\vec{r}_1,\vec{R}-\vec{r}_2)\Phi_H(\vec{r}_3).$$
$$\Psi_2(\vec{r}_1,\vec{r}_2;\vec{r}_3) = \omega_{12}^{[21]}\Phi_{Li^+}(\vec{R}-\vec{r}_1,\vec{R}-\vec{r}_2)\Phi_H(\vec{r}_3) =$$
$$= (P_{23} - P_{13} - P_{123} + P_{132})\Phi_{Li^+}(\vec{R}-\vec{r}_1,\vec{R}-\vec{r}_2)\Phi_H(\vec{r}_3). \quad (56)$$

The related spin parts are

$$X_1 = \omega_{22}^{[21]}\chi(123) = \frac{1}{\sqrt{12}}\{2\chi_{Li^+}(12)\chi_H(3) + \chi_{Li^+}(13)\chi_H(2) - 2\chi_{Li^+}(21)\chi_{e^+}(3) + \chi_{Li^+}(32)\chi_H(1) -$$
$$\chi_{Li^+}(23)\chi_H(1) - \chi_{Li^+}(31)\chi_H(2)\},$$
$$X_2 = \omega_{21}^{[21]}\chi(123) = \frac{1}{2}\{-\chi_{Li^+}(13)\chi_H(2) + \chi_{Li^+}(32)\chi_H(1) - \chi_{Li^+}(23)\chi_H(1) + \chi_{Li^+}(31)\chi_H(2)\}.$$

(57)

The normalisation factor $f_0$ in Eq. 55 is

$$f_0 = \langle \Psi_1(\vec{r}_1,\vec{r}_2,\vec{r}_3)X_1(1,2,3) + \Psi_2(\vec{r}_1,\vec{r}_2,\vec{r}_3)X_2(1,2,3) | \Phi_{Li^+}(\vec{R}-\vec{r}_1,\vec{R}-\vec{r}_2)\Phi_H(\vec{r}_3)\chi(\xi_1,\xi_2;\xi_3), \rangle. \quad (58)$$

By taking into account the orthogonality of the wavefunction spin parts, the normalisation factor $f_0$ is reduced to the expression

$$f_0 = \left\{ \langle \Psi_1(\vec{r}_1,\vec{r}_2,\vec{r}_3) | \Psi_{Li^+}(\vec{r}_1,\vec{r}_2,)\psi_H\vec{r}_3 \rangle \frac{\sqrt{3}}{2} + \langle \Psi_2(\vec{r}_1,\vec{r}_2,\vec{r}_3) | \Psi_{Li^+}(\vec{r}_1,\vec{r}_2,)\psi_H\vec{r} \rangle \frac{1}{2} \right\}. \quad (59)$$

By substituting Eqs. 56 and 57 into Eq. 58, the normalisation factor $f_0$ is found to be

$$f_0 = \int d^3R\, e^{(i(\vec{k}_f-\vec{k}_i)\cdot\vec{R})}\{\langle 2\Phi_{Li^+}(\vec{R}-\vec{r}_1,\vec{R}-\vec{r}_2)\Phi_H(\vec{r}_3) + 2\Phi_{Li^+}(\vec{R}-\vec{r}_2,\vec{R}-\vec{r}_1)\Phi_H(\vec{r}_3) -$$
$$-\Phi_{Li^+}(\vec{R}-\vec{r}_1,\vec{R}-\vec{r}_3)\Phi_H(\vec{r}_2) - \Phi_{Li^+}(\vec{R}-\vec{r}_3,\vec{R}-\vec{r}_1)\Phi_H(\vec{r}_2) - \Phi_{Li^+}(\vec{R}-\vec{r}_2,\vec{R}-\vec{r}_3)\Phi_H(\vec{r}_1) -$$
$$-\Phi_{Li^+}(\vec{R}-\vec{r}_3,\vec{R}-\vec{r}_1)\Phi_H(\vec{r}_2) | \Phi_{Li^+}(\vec{R}-\vec{r}_1,\vec{R}-\vec{r}_2)\Phi_H(\vec{r}_3) \rangle \cdot \frac{\sqrt{3}}{2} +$$
$$\langle \Phi_{Li^+}(\vec{R}-\vec{r}_1,\vec{R}-\vec{r}_3)\Phi_H(\vec{r}_2) - \Phi_{Li^+}(\vec{R}-\vec{r}_3,\vec{R}-\vec{r}_2)\Phi_H(\vec{r}_1) - \Phi_{Li^+}(\vec{R}-\vec{r}_2,\vec{R}-\vec{r}_3)\Phi_H(\vec{r}_1) +$$
$$+\Phi_{Li^+}(\vec{R}-\vec{r}_3,\vec{R}-\vec{r}_1)\Phi_H(\vec{r}_2) | \Phi_{Li^+}(\vec{R}-\vec{r}_1,\vec{R}-\vec{r}_2)\Phi_H(\vec{r}_3) \rangle \frac{1}{2}\}. \quad (60)$$

By substituting Eq. 53 and Eq. 54-a into Eq. 47, we find the differential cross-section for the singlet state:

$$\frac{d\sigma_{if}}{d\Omega} = \frac{\mu_i\mu_f k_f}{(2\pi\hbar^2 |f_0 f_{Li}|)^2 k_i} \times$$
$$\left|\left\langle (\Psi_1\cdot X_1 + \Psi_2\cdot X_2(1,2,3))\exp(i\vec{k}_f\vec{R}) | \hat{T} | (\Psi_{Li1}\cdot X_{Li1} + \Psi_{Li2}\cdot X_{Li2}(1,2,3))\exp(i\vec{k}_i\vec{R}) \right\rangle\right|^2. \quad (61)$$

Here the transition operator $\hat{T}$ is given by Eq. 48, where we used

$$H = H_0^{p=0} + V^{p=0} = H(1,2,3) - \frac{\hbar^2}{2\mu}\nabla_{p-Li}^2 - \frac{e^2}{r_{p1}} - \frac{e^2}{r_{p2}} - \frac{e^2}{r_{p3}} + \frac{3e^2}{R}, \quad (62)$$

and

$$V_0^{\mathbb{N}} = \left(\frac{f_0^2}{P}\right)V_0 = \left(\frac{f_0^2}{6}\right)\left(-\frac{e^2}{r_{p1}} - \frac{e^2}{r_{p2}} - \frac{e^2}{r_{p3}} + \frac{3e^2}{R}\right). \quad (63)$$

The differential cross section given by Eq. 47, becomes in the first approximation

$$\frac{d\sigma_{fi}}{d\Omega} = \frac{\mu_i \mu_f k_f}{(2\pi\hbar^2)^2 k_i} \left| \left( \frac{f_0^2}{6} \right) \langle \Psi_f^0 | V_0 | \Phi_i^0 \rangle \right|^2, \qquad (64)$$

where $\mu_i = \frac{M_{Li} m_p}{(M_{Li} + m_p)} = 1606.79\, m_e$ and $\mu_f = \frac{M_{Li^*} M_H}{(M_{Li^*} + M_H)} = 1607.54\, m_e$ are the reduced mass of the incident particles, the Lithium atom and proton at the beginning of scattering process and the Lithium ion and Hydrogen atom at the end of the-scattering process. The solution to the matrix element in Eq. 64 is provided in Appendix III. Here we would like to emphasize the advantages of applying the EPT method. The matrix element given by Eq. 64 contains the exchange and super-exchange integrals, contained in the second, third and fourth terms in Eq. 64. These integrals take into account the permutations of the electrons between the atoms. The signs of these integrals are defined by the Young diagrams and depend on the total spin value.

The differential cross section given by Eq. 64 is plotted as a function of the scattering angle $\theta$ and $k$ in Fig. 3, where $k$ in the elastic scattering approximation is given by $k = k_i = k_f$. Here, and in what follows, we use Hartree atomic units, where $k$ is measured in reciprocal Bohr radius units ($1/a_B$).

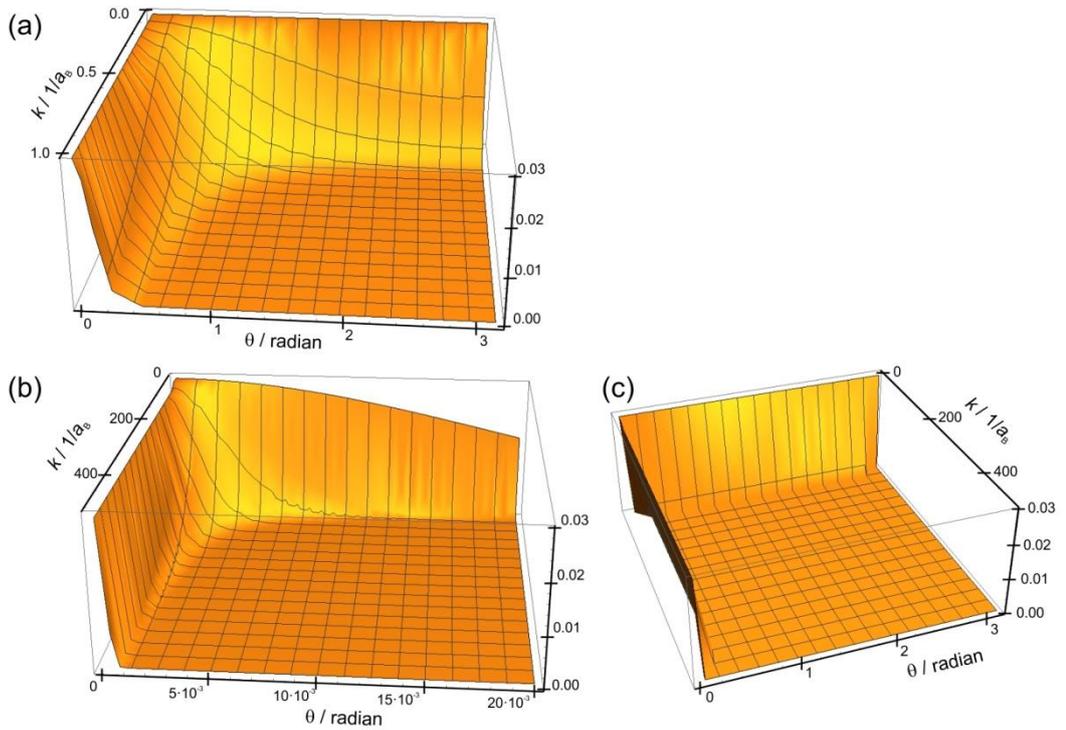

Fig. 3. Simulated differential cross section $\frac{d\sigma_{fi}}{d\Omega}$ [cm² /sr] (a) Differential cross section, for the relative moving energy $E < 0.1$ eV in the range $\theta = [0, \pi]$. (b) The peak of differential cross section observed in the range $k = [0,500]$, corresponding to the relative moving energy $0 < E < 2$ keV and $\theta = [0, 0.2]$. (c) Differential cross section, for the relative moving energy $0 < E < 2$ keV and $\theta = [0, \pi]$.

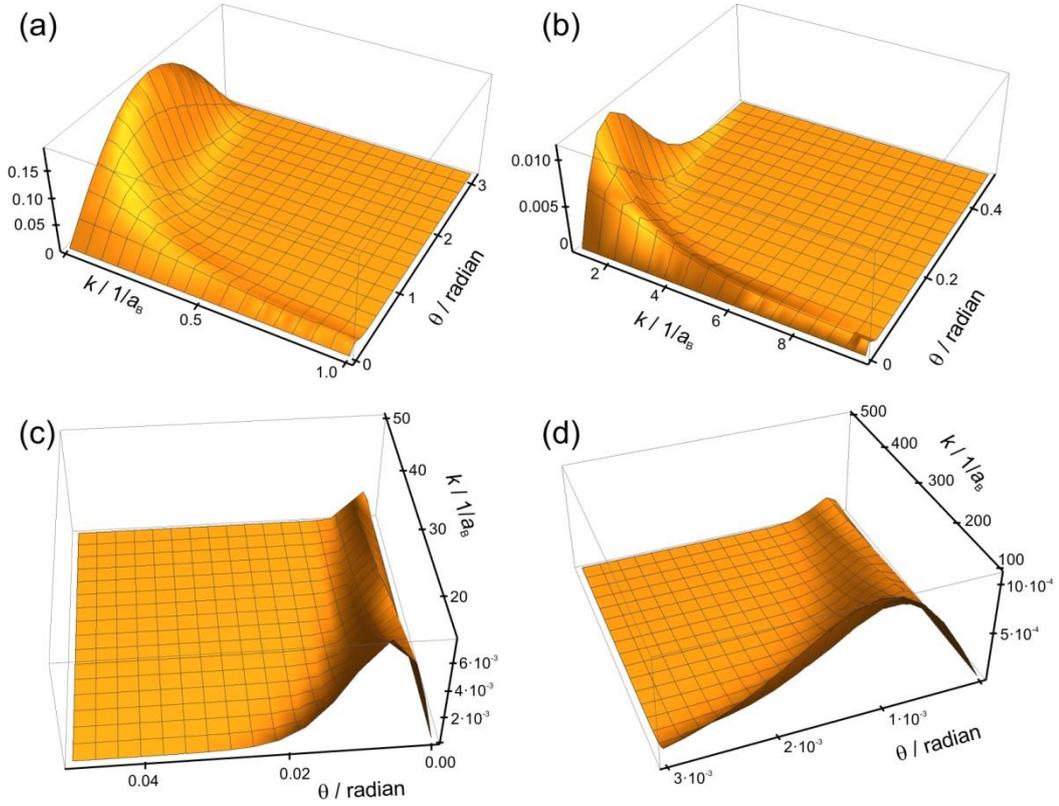

Fig. 4. Simulated differential cross section $\frac{d\sigma_{fi}}{d\Omega} \cdot 2\pi \cdot \sin\theta$ [cm² / rad] at different relative moving energies $E$ as a function of $k$ and $\theta$. (a) $E < 0.01$ eV (i. e. $k < 1.1$ $1/a_B$) and $\theta = [0, \pi]$. (b) $E < 1$ eV (i.e. $k < 10$ $1/a_B$) and $\theta = [0, \pi/6]$. (c) $E < 22$ eV (i.e. $k < 50$ $1/a_B$) and $\theta = [0, \pi/60]$. (d) $25 < E < 2000$ eV (i.e. $50 < k < 500$ $1/a_B$) and $\theta = [0, \pi/100]$.

Fig. 4 shows the simulated $\frac{d\sigma_{fi}}{d\theta} = \int_0^{2\pi} \frac{d\sigma_{fi}}{d\Omega} d\varphi \cdot \sin\theta$ as a function of the scattering angle $\theta$ at different energies of the incident proton. In insets 4(c) and (d), one can observe regions of a "twisted ridge" for certain values of $k$ and $\theta$. It has been previously reported that under similar conditions, but when an alpha-particle is colliding with a Lithium atom [26], the differential cross-section has a smooth appearance without ridges.

To find the total cross section we apply the optical theorem, which states that $\sigma_{fi} = \frac{4\pi}{k} \text{Im} f(\theta)|_{\theta=0}$ [1, 2], where the scattering amplitude $f(\theta)$ is given by $f(\theta) = -\frac{\mu_f}{2\pi\hbar^2} \langle \Psi_f^0 | V_0 | \Phi_i \rangle$:

$$\sigma_{fi} = -\frac{2\mu_f}{k\hbar^2} \text{Im}\left( \langle \Psi_f^0 | V_0 | \Phi_i \rangle \right)_{\theta=0}. \tag{65}$$

The total cross section $\sigma_{fi}$ is plotted in Fig. 5 and also summarized in Table 1, where the values of the cross section obtained in this paper are compared to those obtained from experimental [21,23,24] and theoretical publications [25,26].

In the limit of long wavelengths ($k < 1$), the matrix element in the total cross-section given by Eq. 65 is too large and cannot be simplified to a short expression. Therefore, first, the differential cross-section for $k < 1$ was approximated by

$$\frac{d\sigma_{fi}}{d\Omega}(k<1) = \frac{1.7 \cdot 10^{-10}}{0.25 \cdot 10^{-16}} \pi \cdot \left| \frac{-9 + i\left(62.5k \cdot \sin(\theta/2) - 144(k \cdot \sin(\theta/2))^3\right)}{1.6^2(0.6 + 4k^2)^2} + 10 \right|^2 =$$
$$= 6.8 \cdot 10^6 \pi \cdot \left| \frac{-9 + i\left(62.5k \cdot \sin(\theta/2) - 144(k \cdot \sin(\theta/2))^3\right)}{1.6^2(0.6 + 4k^2)^2} + 10 \right|^2. \tag{66}$$

and then it was integrated over $\theta$ and $\varphi$. The result amounts to

$$\sigma_{fi}(k<1) = 6.8 \cdot \pi \cdot \frac{\left(7812.5k^2 - 18000k^5 + 20736k^6 + 4\left(6.36 + 102.4k^2\right)^2\right)}{\left(0.6 + 4k^2\right)^2}, \tag{67}$$

and is plotted in Fig. 5(a).

In the limit of short wavelengths ($k > 5$), the imaginary part of the scattering amplitude, as follows from Eq. 62, amounts to

$$\text{Im}[f(\theta)] = \frac{\mu_f}{2\pi\hbar^2} \cdot \frac{10.8}{2\pi^3} \left\{ \frac{\left(625 + 144k^2 \sin^2\left[\frac{\theta}{2}\right]\right)\left(16k^2 \sin^2\left[\frac{\theta}{2}\right] - 2.8^2\right) + 151 \cdot 22.4k^2 \sin^2\left[\frac{\theta}{2}\right]}{\left(16k^2 \sin^2\left[\frac{\theta}{2}\right] + 2.8^2\right)} + \right.$$
$$\left. + \frac{0.9\left(\left(2k\sin\left[\frac{\theta}{2}\right]\right)^3 - 486k\sin\left[\frac{\theta}{2}\right]\right)}{\left(0.81 + 4k^2 \sin^2\left[\frac{\theta}{2}\right]\right)^3} \right\}. \tag{68}$$

This gives the expression for the total cross section in atomic units, that is in Bohr radius squared

$$\sigma_{fi}(k > 5) = \frac{4\pi}{k} \operatorname{Im} f(\theta)\big|_{\theta=0} = \frac{1}{k} \frac{\mu_f}{\hbar^2} \cdot \frac{10.8}{\pi^3} \cdot 625 \tag{69}$$

The total cross section for short wavelengths ($k > 5$) given by Eq. 69 is plotted in Fig. 5 (b)–(e).

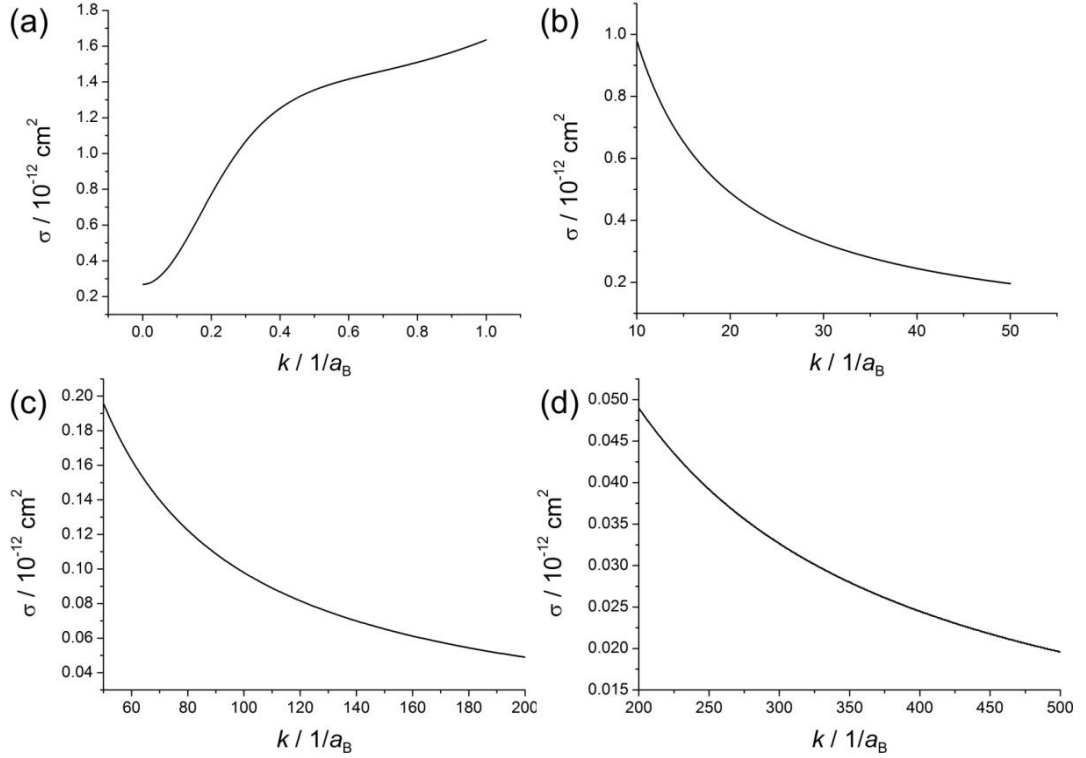

Fig. 5. Total cross section $\sigma_{fi}$ of proton scattering on a Lithium atom for the incident energy ranging: (a) $E < 0.01$ eV ($k <1$ $1/a_B$), (b) $E = 0.087 – 21.6$ eV ($k = 10.–50$ $1/a_B$), (c) $E = 21.6 – 346$ eV ($k = 50 – 200$ $1/a_B$), (d) $E = 346 – 2161$ eV ($k = 200 – 500$ $1/a_B$).

| $v$ (cm/s) | $E$ (eV) | $k$ ($1/a_B$) | $\sigma$ (cm$^2$) Li – H$^+$ as calculated from Eq.62 and Eq.64 | $\sigma$ (cm$^2$) Li –He$^{2+}$ [26] | $\sigma$ (cm$^2$) Li – Li$^+$ He – He$^+$ H – H$^+$ [25] | $\sigma$(cm$^2$) He – He$^+$ [21, 23, 24] |
|---|---|---|---|---|---|---|
| $2.5 \cdot 10^5$ | 0.22 | 5 | $2.0 \cdot 10^{-13}$ | $5 \cdot 10^{-13}$ | | |
| $5 \cdot 10^5$ | 0.87 | 10 | $1.5 \cdot 10^{-13}$ | $1.30 \cdot 10^{-13}$ | | |
| $7.5 \cdot 10^5$ | 1.95 | 15 | $7.5 \cdot 10^{-14}$ | $5.8 \cdot 10^{-14}$ | $2.6 \cdot 10^{-14}$ $3.5 \cdot 10^{-15}$ $6.2 \cdot 10^{-15}$ | |
| $1 \cdot 10^6$ | 3.45 | 20 | $5.3 \cdot 10^{-14}$ | $32.5 \cdot 10^{-15}$ | | |
| $1.25 \cdot 10^6$ | 5.4 | 25 | $4.5 \cdot 10^{-14}$ | $26.4 \cdot 10^{-15}$ | | |
| $2.5 \cdot 10^6$ | 21.6 | 50 | $2.3 \cdot 10^{-14}$ | $17 \cdot 10^{-15}$ | $2.2 \cdot 10^{-14}$ $2.8 \cdot 10^{-15}$ $5 \cdot 10^{-15}$ | |
| $5 \cdot 10^6$ | 86.4 | 100 | $1.0 \cdot 10^{-14}$ | $10 \cdot 10^{-15}$ | | |
| $7.5 \cdot 10^6$ | 195 | 150 | $7.0 \cdot 10^{-15}$ | $8.75 \cdot 10^{-15}$ | $18 \cdot 10^{-15}$ $2.1 \cdot 10^{-15}$ $3.8 \cdot 10^{-15}$ | |
| $1 \cdot 10^7$ | 345 | 200 | $6.5 \cdot 10^{-15}$ | $8 \cdot 10^{-15}$ | | |
| $2.5 \cdot 10^7$ | 2161 | 500 | $2.5 \cdot 10^{-15}$ | $5.6 \cdot 10^{-15}$ | | |
| $5 \cdot 10^7$ | 8646 | 1000 | $1.0 \cdot 10^{-15}$ | $5.32 \cdot 10^{-15}$ | | |
| $1 \cdot 10^8$ | 3458 | 2000 | $6.5 \cdot 10^{-16}$ | $5.24 \cdot 10^{-15}$ | | $5 \cdot 10^{-16}$ |
| $1.5 \cdot 10^8$ | 77821 | 3000 | $4 \cdot 10^{-16}$ | $5.22 \cdot 10^{-15}$ | | $3.5 \cdot 10^{-16}$ |
| $2 \cdot 10^8$ | 138348 | 4000 | $3.6 \cdot 10^{-16}$ | $5.21 \cdot 10^{-15}$ | | $2.7 \cdot 10^{-16}$ |
| $2.5 \cdot 10^8$ | 216169 | 5000 | $3 \cdot 10^{-16}$ | $5.21 \cdot 10^{-15}$ | | $2 \cdot 10^{-16}$ |
| $3 \cdot 10^8$ | | 6000 | $2 \cdot 10^{-16}$ | $5.20 \cdot 10^{-15}$ | | $1 \cdot 10^{-16}$ |

Table 1. Summary of total cross section as a function of the relative collision energy $E$. Here, $v$ (cm/s) is the speed of the relative motion. The relative collision energy is calculated as $E = \frac{k^2 \hbar^2}{2\mu_i}$ and the relative collision speed is calculated as $v = \frac{\hbar k}{\mu_i}$.

## Conclusions

The formalism of the time-dependent exchange perturbation theory is developed in an invariant form. It allows the scattering processes of complex particles (atoms, molecules) to be described, taking into account the indistinguishability of electrons participating in multi-center collisions, even in cases where the permutations of electrons occur between different centers and are associated with non-orthogonal states. This formalism is applicable to cases of restructuring colliding particles, such as ion charge exchange processes.

**Appendix I**

We apply $\hat{O}_i = 1 - |\Psi_i^0\rangle\langle\Phi_i|$ to both sides of equation Eq. 12:

$$-\frac{\hbar}{i}|\dot{\Psi}^{(1)}(t)\rangle = \hat{H}_0|\Psi^{(1)}(t)\rangle + \hat{V}(t)|\Psi_i^0\rangle \tag{A1}$$

where we use the expansion given by Eq. 13:

$$|\Psi^{(1)}(t)\rangle = \sum_n C_n(t)\exp\left(-\frac{i}{\hbar}E_n t\right)|\Psi_n^0\rangle \tag{A2}$$

The resulting equation can now be written as

$$-\frac{\hbar}{i}\sum_{n}{}'\dot{C}_n(t)\exp\left(-\frac{i}{\hbar}E_n t\right)\hat{O}_i\left|\Psi_n^0\right\rangle = \exp\left(-\frac{i}{\hbar}E_i t\right)\hat{O}_i\hat{V}\left|\Psi_i^0\right\rangle, \quad (A3)$$

where prime denotes that the term with $n=i$ is excluded.

Next, we use the completeness property of the antisymmetrised basis (see Eq. 4) and re-write the right side of Eq. A3 as

$$\exp\left(-\frac{i}{\hbar}E_i t\right)\cdot\frac{f_0}{P}\sum_{n}{}'\left|\Psi_n^0\right\rangle\left(\Phi_n^{0(0)}\left|\hat{O}_i\hat{V}\right|\Psi_i^0\right) \quad (A4)$$

Because $\left(\Phi_i^{0(0)}\left|\hat{O}_i\right. = \left(\Phi_i^{0(0)}\right| - \left(\Phi_i^{0(0)}\left|\Psi_i^0\right)\left(\Phi_i^{0(0)}\right|\right. \equiv 0\right.$, the term at $n=i$, is excluded, which is signified by the prime on the summation sign.

Rearranging Eq. A3, we obtain

$$\sum_{n}{}'\left\{-\frac{\hbar}{i}\dot{C}_n(t) - \frac{f_0}{P}\left(\Phi_n^{0(0)}\left|\hat{O}_i\hat{V}\right|\Psi_i^0\right)\exp(-i\omega_{in}t)\right\}\hat{O}_i\left|\Psi_n^0\right\rangle = 0. \quad (A5)$$

The last equation is fulfilled when the expression in the curly brackets equals zero for all $n$. This, in turn, leads to an equation for $\dot{C}_n(t)$:

$$-\frac{\hbar}{i}\dot{C}_n(t) = \frac{f_0}{P}\exp(-i\omega_{ni}t)\left(\Phi_n^{0(0)}\left|\hat{O}_i\hat{V}\right|\Psi_i^0\right), \quad (A6)$$

where we introduced $\frac{1}{\hbar}(E_i - E_n) = \omega_{in}$.

**Appendix II**

Operator $\hat{T}$ satisfies Eq. 48:

$$\hat{T} = V_0^{\mathbb{N}} + V_0^{\mathbb{N}}\left(\frac{f_0^2}{P}\right)^{-1}\left(E_i - H_{p=0}^0 + i\eta\right)^{-1}\hat{T}, \quad (A7)$$

which can be re-arranged to:

$$V_0^{\mathbb{N}} = (\hat{1} - V_0^{\mathbb{N}}\left(\frac{f_0^2}{P}\right)^{-1}\left(E_i - H_{p=0}^0 + i\eta\right)^{-1})\hat{T}$$

$$V_0^{\mathbb{N}} = \left(\left(E_i - H_{p=0}^0 + i\eta\right)\left(\frac{f_0^2}{P}\right) - V_0^{\mathbb{N}}\right)\left(\frac{f_0^2}{P}\right)^{-1}\left(E_i - H_{p=0}^0 + i\eta\right)^{-1}\hat{T}. \quad (A8)$$

Taking into account that the total Hamiltonian of the system

$$H = H^0_{p=0} + V_{p=0} = H^0_p + V_p \quad , \tag{A9}$$

is invariant with respect to permutations of electrons between atoms:

$$H = H_0^{p=0} + V^{p=0} = -\frac{\hbar^2}{2\mu_i}\nabla_i^2 + H_i^{p=0}(r_1, r_2, \ldots)_i + V^{p=0}. \tag{A10}$$

We further re-arrange Eq. A8:

$$V_0^{\mathbb{N}} = (E_i - H + i\eta)\left(\frac{f_0^2}{P}\right)\left(\frac{f_0^2}{P}\right)^{-1}(E_i - H^0_{p=0} + i\eta)^{-1}\hat{T}$$

$$\left(\frac{f_0^2}{P}\right)(E_i - H^0_{p=0} + i\eta)\left(\frac{f_0^2}{P}\right)^{-1}(E_i - H + i\eta)^{-1}V_0^{\mathbb{N}} = \hat{T}$$

$$\left(\frac{f_0^2}{P}\right)(E_i - H^0_{p=0} + V_0 - V_0 + i\eta)\left(\frac{f_0^2}{P}\right)^{-1}(E_i - H + i\eta)^{-1}V_0^{\mathbb{N}} = \hat{T}$$

$$V_0^{\mathbb{N}}\left(\frac{f_0^2}{P}\right)^{-1}(E_i - H + i\eta)^{-1}V_0^{\mathbb{N}} + \left(\frac{f_0^2}{P}\right)(E_i - H^0_{p=0} - V_0 + i\eta)\left(\frac{f_0^2}{P}\right)^{-1}(E_i - H + i\eta)^{-1}V_0^{\mathbb{N}} =$$

$$= \hat{T}$$

to the final expression for operator $\hat{T}$:

$$V_0^{\mathbb{N}}\left(\frac{f_0^2}{P}\right)^{-1}(E_i - H + i\eta)^{-1}V_0^{\mathbb{N}} + V_0^{\mathbb{N}} = \hat{T}. \tag{A11}$$

**Appendix III**

The initial states of the electrons in the Lithium atom are given by [27] where the parameters α and β are taken from [28]

$$\phi_{1s}(\vec{R} - \vec{r}_i) = (\alpha_1^3/\pi)^{1/2}\exp(-\alpha_1|\vec{R} - \vec{r}_i|), \quad i = 1, 2$$
$$\varphi_{2s^1}(\vec{R} - \vec{r}_3) = (\alpha_2^3/8\pi)^{1/2}(1 - 0.5\alpha_2|\vec{R} - \vec{r}_3|)\exp(-0.5\alpha_2|\vec{R} - \vec{r}_3|), \tag{A12}$$
$$\text{where} \quad \alpha_1 = 2.698, \alpha_2 = 0.795.$$

The final states of the electrons in the Helium-like Lithium ion are described by

$$\phi^*(\vec{R} - \vec{r}_i) = (\alpha^{*3}/\pi)^{1/2}\exp(-\alpha^*|\vec{R} - \vec{r}_i|), \quad i = 1, 2$$
$$\text{where} \quad \alpha^* = 1.692. \tag{A13}$$

The final state of the electron in the Hydrogen atom is described by the single-electron wave function:

$$\psi_H(r) = (\pi)^{-1/2} \exp(-\beta r);$$
$$\beta = 1$$
(A14)

Here α, α* and β are given in reciprocal Bohr radius units, and $R$ is in Bohr radius units.

The first-order correction to the S-matrix elements given by Eq. 31 becomes:

$$\langle \Psi_f^0 | V_0^{\mathbb{N}} | \Phi_i \rangle =$$
$$= \frac{1}{f_0} \frac{1}{f_{Li}} \left( \sqrt{3} \langle \Psi_1(\vec{r}_1,\vec{r}_2,\vec{r}_3) | V_0^{\mathbb{N}} e^{(i(\vec{k}_f - \vec{k}_i)\cdot \vec{R}_{I,II})} | \Psi_{Li1}(\vec{r}_1,\vec{r}_2,\vec{r}_3) \rangle + \frac{3}{2} \langle \Psi_2(\vec{r}_1,\vec{r}_2,\vec{r}_3) | V_0^{\mathbb{N}} e^{(i(\vec{k}_f - \vec{k}_i)\cdot \vec{R}_{I,II})} | \Psi_{Li2}(\vec{r}_1,\vec{r}_2,\vec{r}_3) \rangle \right) =$$
$$= \frac{4}{\sqrt{3}} \frac{1}{P} \frac{f_0}{f_{Li}} \Big\{ \langle \Phi_{Li^+}(\vec{R}-\vec{r}_1,\vec{R}-\vec{r}_2) \Phi_H(\vec{r}_3) | V_0 e^{(i(\vec{k}_f - \vec{k}_i)\cdot \vec{R}_{I,II})} | (\phi_{1s}(\vec{R}-\vec{r}_1)\phi_{1s}(\vec{R}-\vec{r}_2)\varphi_{2s}(\vec{R}-\vec{r}_3)) \rangle -$$
$$- \langle \Phi_{Li^+}(\vec{R}-\vec{r}_1,\vec{R}-\vec{r}_2) \Phi_H(\vec{r}_3) | V_0 e^{(i(\vec{k}_f - \vec{k}_i)\cdot \vec{R}_{I,II})} | \phi_{1s}(\vec{R}-\vec{r}_3)\phi_{1s}(\vec{R}-\vec{r}_2)\varphi_{2s}(\vec{R}-\vec{r}_1) \rangle -$$
$$- \langle \Phi_{Li^+}(\vec{R}-\vec{r}_3,\vec{R}-\vec{r}_2) \Phi_H(\vec{r}_1) | V_0 e^{(i(\vec{k}_f - \vec{k}_i)\cdot \vec{R}_{I,II})} | (\phi_{1s}(\vec{R}-\vec{r}_1)\phi_{1s}(\vec{R}-\vec{r}_2)\varphi_{2s}(\vec{R}-\vec{r}_3)) \rangle +$$
$$+ \langle \Phi_{Li^+}(\vec{R}-\vec{r}_3,\vec{R}-\vec{r}_2) \Phi_H(\vec{r}_1) | V_0 e^{(i(\vec{k}_f - \vec{k}_i)\cdot \vec{R}_{I,II})} | \phi_{1s}(\vec{R}-\vec{r}_3)\phi_{1s}(\vec{R}-\vec{r}_2)\varphi_{2s}(\vec{R}-\vec{r}_1)) \rangle \Big\}$$
(A15)

where we used the orthogonality of the spin parts of the wavefunctions.

The matrix element $\langle \Psi_f^0 | V_0 | \Phi_i \rangle$ in Eq. 60 can be re-written as

$$\langle \Psi_f^0 | V_0 | \Phi_i \rangle = \frac{4}{\sqrt{3}} \frac{1}{P} \frac{f_0}{f_{Li}} \int d^3 R\, e^{(i(\vec{k}_f - \vec{k}_i)\cdot \vec{R})} \left\{ \Delta_{1s*1s} \left[ \left( \frac{6}{R} \Delta_{1s*1s} S_{1s'2s} - 2A_{1s'2s}\Delta_{1s*1s} - 4K_{1s*1s} S_{1s'2s} \right) - \right. \right.$$
$$\left. - \left( S_{1s'2s}(\frac{3}{R}\Delta_{1s*1s} - K_{1s*1s} - K_{1s*2s}) - A_{1s'1s}\Delta_{1s*2s} \right) \right] -$$
$$- \Delta_{1s*1s}\Delta_{1s*2s}\left( \frac{3S_{1s'1s}}{R} - A_{1s'1s} \right) + S_{1s'1s}\left( K_{1s*1s}\Delta_{1s*2s} + K_{1s*2s}\Delta_{1s*1s} \right) \Big\},$$
(A16)

where we introduced the following symbols for integrals

$$\begin{aligned}
\Delta_{1s*1s} &= \langle \phi^* | \phi_{1s^1} \rangle & K_{1s*1s} &= \langle \phi^* | \frac{1}{r} | \phi_{1s^1} \rangle \\
\Delta_{1s*2s} &= \langle \phi^* | \phi_{2s^1} \rangle & K_{1s*2s} &= \langle \phi^* | \frac{1}{r} | \phi_{2s^1} \rangle \\
S_{1s'1s} &= \langle \psi_{He^*}(\vec{r}) | \phi_{1s}(\vec{R}-\vec{r}) \rangle & A_{1s'1s} &= \langle \psi_{He^*}(\vec{r}) | \frac{1}{r} | \phi_{1s}(\vec{R}-\vec{r}) \rangle \\
S_{1s'2s} &= \langle \psi_{He^*}(\vec{r}) | \phi_{2s}(\vec{R}-\vec{r}) \rangle & A_{1s'2s} &= A_{1s'1s} = \langle \psi_{He^*}(\vec{r}) | \frac{1}{r} | \phi_{2s}(\vec{R}-\vec{r}) \rangle .
\end{aligned}$$
(A17)

All these integrals have analytical expressions, and they are listed below.

The normalization factor in Eq. 60 is found as follows:

$$f_0 = \frac{4}{3} \times$$

$$\langle 4\phi^*(\vec{R}-\vec{r}_1)\cdot\phi^*(\vec{R}-\vec{r}_2)\cdot\psi_{He^*}(r_3) - 3\phi^*(\vec{R}-\vec{r}_1)\cdot\phi^*(\vec{R}-\vec{r}_3)\cdot\psi_{He^*}(r_2) - \phi^*(\vec{R}-\vec{r}_3)\cdot\phi^*(\vec{R}-\vec{r}_2)\cdot\psi_{He^*}(r_1) |$$

$$| e^{(i(\vec{k}_f - \vec{k}_i)\cdot\vec{R})} (\phi_{1s}(\vec{R}-\vec{r}_1)\phi_{1s}(\vec{R}-\vec{r}_2)\varphi_{2s}(\vec{R}-\vec{r}_3) - \phi_{1s}(\vec{R}-\vec{r}_3)\phi_{1s}(\vec{R}-\vec{r}_2)\varphi_{2s}(\vec{R}-\vec{r}_1)) \rangle =$$

$$= \frac{20}{3}\int d^3R\, e^{(i(\vec{k}_f - \vec{k}_i)\cdot\vec{R})} \{\Delta_{1s^*1s}\Delta_{1s^*1s}S_{1s'2s} - \Delta_{1s^*1s}\Delta_{1s^*2s}S_{1s'1s}\}$$

(A18)

$$\Delta_{1s^*1s} = \langle\phi^*|\phi_{1s^1}\rangle = \frac{8(\alpha\alpha^*)^{3/2}}{(\alpha+\alpha^*)^3}$$

$$\Delta_{1s^*2s} = \langle\phi^*|\phi_{2s^1}\rangle = \frac{8\left(\frac{\alpha}{2}\alpha^*\right)^{3/2}}{\left(\frac{\alpha}{2}+\alpha^*\right)^3}\left(1 - \frac{3\alpha}{\alpha+2\alpha^*}\right)$$

(A19)

$$S_{1s'1s} = \langle\psi_{He^*}(\vec{r})|\phi_{1s}(\vec{R}-\vec{r})\rangle =$$

$$= \frac{8(\alpha\beta)^{3/2}}{\beta^2-\alpha^2}\mathrm{sh}\left[\frac{R}{2}(\alpha-\beta)\right]e^{-\frac{R}{2}(\alpha+\beta)}\left(\frac{1}{\alpha+\beta} + \frac{\mathrm{cth}\frac{R}{2}(\alpha-\beta)}{\alpha-\beta} - \frac{8\alpha\beta}{R(\alpha^2-\beta^2)^2}\right)$$

$$S_{1s'2s} = \langle\psi_{He^*}(\vec{r})|\phi_{2s}(\vec{R}-\vec{r})\rangle = \frac{(\alpha\beta)^{3/2}}{\sqrt{2}}\frac{R\cdot\mathrm{sh}\left[\frac{R}{2}(\frac{\alpha}{2}-\beta)\right]e^{-\frac{R}{2}(\frac{\alpha}{2}+\beta)}}{\beta^2-\left(\frac{\alpha}{2}\right)^2}\times$$

$$\left\{\frac{4}{R\left(\frac{\alpha}{2}+\beta\right)}\frac{8\alpha\beta}{R^2\left(\left(\frac{\alpha}{2}\right)^2-\beta^2\right)^2} + \frac{4\mathrm{cth}\left[\frac{R}{2}(\frac{\alpha}{2}-\beta)\right]}{R\left(\frac{\alpha}{2}-\beta\right)} - \frac{\alpha\left(1-\mathrm{cth}\left[\frac{R}{2}(\frac{\alpha}{2}-\beta)\right]\right)}{R\left(\frac{\alpha}{2}+\beta\right)} - \right.$$

$$-\frac{2\alpha\left(3-\mathrm{cth}\left[\frac{R}{2}(\frac{\alpha}{2}-\beta)\right]\right)}{R\left(\frac{\alpha}{2}+\beta\right)^2} - \frac{12\alpha}{R^2\left(\left(\frac{\alpha}{2}\right)+\beta\right)^3} + \frac{12\alpha}{R^2\left(\left(\frac{\alpha}{2}\right)-\beta\right)^3} +$$

$$\left. + \frac{\alpha\left(1-\mathrm{cth}\left[\frac{R}{2}(\frac{\alpha}{2}-\beta)\right]\right)}{R\left(\frac{\alpha}{2}-\beta\right)} - \frac{2\alpha\left(1+\mathrm{cth}\left[\frac{R}{2}(\frac{\alpha}{2}-\beta)\right]\right)}{R\left(\left(\frac{\alpha}{2}\right)^2-\beta^2\right)} + \frac{2\alpha\left(1-3\mathrm{cth}\left[\frac{R}{2}(\frac{\alpha}{2}-\beta)\right]\right)}{R\left(\frac{\alpha}{2}-\beta\right)^2}\right\}$$

(A20)
$$K_{1s*1s} = \langle \phi^* | \frac{1}{r} | \phi_{1s^1} \rangle = \frac{4(\alpha\alpha^*)^{3/2}}{(\alpha+\alpha^*)^2} \text{sh}\left[\frac{R}{2}(\alpha+\alpha^*)\right] e^{-\frac{R}{2}(\alpha+\alpha^*)} \left(1 + \frac{4}{R(\alpha+\alpha^*)} - \text{cth}\left[\frac{R}{2}(\alpha+\alpha^*)\right]\right)$$

(A21)
$$A_{1s'1s} = \langle \psi_{He^*}(\vec{r}) | \frac{1}{r} | \phi_{1s}(\vec{R}-\vec{r}) \rangle = \frac{4(\alpha\beta)^{3/2}}{(\alpha^2-\beta^2)} \text{sh}\left[\frac{R}{2}(\alpha-\beta)\right] e^{-\frac{R}{2}(\alpha+\beta)} \left(1 + \frac{4\alpha}{R(\alpha^2-\beta^2)} - \text{cth}\left[\frac{R}{2}(\alpha-\beta)\right]\right)$$

$$K_{1s*2s} = \langle \phi^* | \frac{1}{r} | \phi_{2s^1} \rangle = \frac{\sqrt{2}(\alpha\alpha^*)^{3/2}}{\left(\frac{\alpha}{2}+\alpha^*\right)^2} \text{sh}\left[\frac{R}{2}\left(\frac{\alpha}{2}+\alpha^*\right)\right] e^{-\frac{R}{2}\left(\frac{\alpha}{2}+\alpha^*\right)} \left(1 + \frac{4}{R\left(\frac{\alpha}{2}+\alpha^*\right)} - \text{cth}\left[\frac{R}{2}\left(\frac{\alpha}{2}+\alpha^*\right)\right] - \right.$$

$$\left. -\frac{R\alpha}{2}\left(1-\text{cth}\left[\frac{R}{2}\left(\frac{\alpha}{2}+\alpha^*\right)\right]\right) - \frac{2\alpha\left(1-\text{cth}\left[\frac{R}{2}\left(\frac{\alpha}{2}+\alpha^*\right)\right]\right)}{\left(\frac{\alpha}{2}+\alpha^*\right)} - \frac{6\alpha}{R\left(\frac{\alpha}{2}+\alpha^*\right)^2}\right)$$

$$A_{1s'2s} = A_{1s'1s} = \langle \psi_{He^*}(\vec{r}) | \frac{1}{r} | \phi_{2s}(\vec{R}-\vec{r}) \rangle = \frac{\sqrt{2}(\alpha\beta)^{3/2}}{\left(\frac{\alpha}{2}\right)^2-\beta^2} \text{sh}\left[\frac{R}{2}\left(\frac{\alpha}{2}-\beta\right)\right] e^{-\frac{R}{2}\left(\frac{\alpha}{2}+\beta\right)} \times$$

$$\left\{1-\text{cth}\left[\frac{R}{2}\left(\frac{\alpha}{2}-\beta\right)\right] - \frac{R\alpha}{2}\left(1-\text{cth}\left[\frac{R}{2}\left(\frac{\alpha}{2}-\beta\right)\right]\right) - \frac{\alpha\left(1-\text{cth}\left[\frac{R}{2}\left(\frac{\alpha}{2}-\beta\right)\right]\right)}{\left(\frac{\alpha}{2}+\beta\right)} - \right.$$

$$\left. -\frac{\alpha\left(1-\text{cth}\left[\frac{R}{2}\left(\frac{\alpha}{2}-\beta\right)\right]\right)}{\left(\frac{\alpha}{2}-\beta\right)} - \frac{2\alpha}{R\left(\frac{\alpha}{2}+\beta\right)^2} - \frac{2\alpha}{R\left(\frac{\alpha}{2}-\beta\right)^2}\right\}$$

(A22)